# Surface-charge-mediated Formation of H-TiO$_2$@Ni(OH)$_2$ Heterostructures for High-Performance Supercapcitors

*Qingqing Ke, Cao Guan, Xiao Zhang, Minrui Zheng, Yong-Wei Zhang, Yongqing Cai,\* Hua Zhang,\* John Wang\**

Dr. Qingqing Ke,
Institute of Materials Research and Engineering, A\*STAR, Singapore 138634;
Department of Materials Science and Engineering, National University of Singapore, 117574 Singapore
Dr. Cao Guan, Prof. Dr. John Wang
Department of Materials Science and Engineering, National University of Singapore, 117574 Singapore
E-mail: msewangj@nus.edu.sg
Dr. Xiao Zhang,
Prof. Dr. Hua Zhang
Center for Programmable Materials, School of Materials Science and Engineering, Nanyang Technological University, 50 Nanyang Avenue, 639798 Singapore
E-mail: hzhang@ntu.edu.sg
Dr. Minrui Zhang
Department of Physics, National University of Singapore, 2 Science Drive 3, 117542, Singapore
Prof. Dr. Yong-Wei Zhang,  Dr. Yongqing Cai
Institute of High Performance Computing, A\*STAR, 138732 Singapore
E-mail: caiy@ihpc.a-star.edu.sg

Supercapacitors or ultracapacitors are promising for efficient energy storage applications, owing to their high power density, high charge-discharge rates, and long cycle life performance.[1] To achieve this goal, a large specific surface area, an high electronic conductivity and a fast cation intercalation/de-intercalation process are generally required in the design and preparation of materials for high-performance supercapacitors.[2] Recently, core-shell heterostructures with multifunctionalities are regarded as one of promising materials for supercapacitors or ultracapacitors applications.[3] In particular, one-dimensional (1D) core-shell heterostructures have sparked great scientific and technological interests due to their high versatility and applicability as the essential components in nanoscale electronics, catalysis, chemical sensing, and energy conversion storage devices.[4,5] Various metal@metal oxide, metal@metal, metal oxide@metal oxide and metal oxide@conductive polymers so far have been investigated.[3] Transition metal hydroxide/oxide $Co_3O_4$, $Co(OH)_2$, $MnO_2$, $Mn(OH)_2$, NiO, $Ni(OH)_2$ and their compounds storing energy by surface faradaic (redox) reactions were generally integrated with conducting scaffold to build core-shell structure.[6-8] Intensive studies show that an enlarged active surface area of transition metal hydroxide/oxide enables a promoted surface redox reaction and enhanced electrochemical performance. Therefore, controllable synthesis of a hierarchically porous construction with high surface areas is critically important for energy storage.

Interface engineering is now proven as an effective way to improve the electrochemical performance and structural integrity of electrode materials.[8-10] In the heterostructures, the presence of interface between the core and shell layers exhibits charge discontinuity and built-in electric field which impose significant effects on the charge flow and accordingly the surface faradaic (redox) reactions. For instance, through tuning the phase across interface from amorphous to crystallized state, a fast transfer of ions and electrons was demonstrated.[11] Significantly, morphologies or the structure features of core-shell hybrid can also be rationalized through modifying the interface/surface conditions. Liu et al., engineered the core-shell ($Co_3O_4$@$MnO_2$) interface by introducing a thin coating layer of carbon, which confines the $MnO_2$ growth reaction specifically to $Co_3O_4$ nanowires (NWs) surface, giving rise to well-constructed hybrid architectures.[12] Hou et al., functionalized the CNTs surfaces by acid treatment to attach carboxylic groups or hydroxyl groups and induce the formation of the hierarchical $MnO_2$ nanophere on CNTs network.[13] Despite the above efforts, fabrication of properly engineered core-shell NWs arrays with well-defined morphologies and tunable functions still remains a challenge. Identifying the underlying mechanism for improved electrochemical performance and seeking ways to optimize the shell configuration through interface engineering are thus highly desired.

In this work, we demonstrate a surface-charge-mediated way to rationalize the core-shell nanostructures, based on judiciously functionalizing the backbone surface with introducing defects on $TiO_2$ core NWs through hydrogenation, together with the subsequently coating $Ni(OH)_2$ shell. It shows that the hydrogenation procedure modulates the surface chemical activity of $TiO_2$ core material and induces the formation of an electrochemically favourable porous $Ni(OH)_2$ with ultrathin nanosheets. Our experimental and theoretical (first-princples) calculations) results reveal that the surface-charge-mediated synthesis by hydrogenation treatment is feasible to fabricate a high-performance supercapcitor due to the following dual effects: (i) hydrogenation boosts the electrical conductivity of $TiO_2$ by three orders of magnitudes, which is beneficial for the high-density electrochemical performance, (ii) the introduction of charged defects, such as oxygen vacancy, or $Ti^{3+}$/$Ti^{4+}$ are conceived to mediate the growth of the $Ni(OH)_2$. Moreover, an asymmetric supercapacitor, with N-doped carbon NWs on carbon cloth (N-C NWs) as the negative electrode, is fabricated with a superior performance in terms of both energy storage capabiltiy and mechanical flexibiltity.

The growth procedure of the core-shell NW electrodes is illustrated in **Figure 1a**. $TiO_2$ NWs were first prepared on a carbon cloth by hydrothermal method. Subsequently, $H_2$ plasma was employed to irradiate the TiO$_2$ NWs surfaces to generate the hydrogenated TiO$_2$ NWs (H-TiO$_2$ NWs). Ni(OH)$_2$ nanoflakes were finally coated onto the H-TiO$_2$ NWs surface by a chemical bath deposition (CBD) process (see the Experimental Section). The H-TiO$_2$ NWs (inset of Figure 1b) show a darker color compared to the white image of pristine TiO$_2$ (Figure S1). Scanning electron microscopy (SEM) image (Figure 1b) shows that H-TiO$_2$ NWs in diameters of ~300 nm are vertically grown on the carbon fibers without destroying the NW architecture. XRD results indicate the formation of a rutile phase of H-TiO$_2$ and β-Ni(OH)$_2$ (Figure S2). Figure 1c and Figure S3 demonstrate the interconnected Ni(OH)$_2$ nanoflakes fully cover the scaffold of H-TiO$_2$ NWs, as indication of a core-shell configuration. Compared to the Ni(OH)$_2$ nanoflakes grown on TiO$_2$, Ni(OH)$_2$ on the H-TiO$_2$ NWs shows a more porous structure with thinner sheets. Such rationalized configuration is able to provide both large active surface areas and desired electrical connection for fast redox kinetics, thus potentially boosting the efficiency of the energy storage.[6] The core-shell H-TiO$_2$@Ni(OH)$_2$ phases were further elucidated by transmission electron microcopy (TEM) and selected area electron diffraction (SAED) analysis. Figure 1d shows an atomic scale profile of the H-TiO$_2$ NW crystal structure by the HRTEM image. The adjacent spaces of around 0.25 and 0.29 nm in H-TiO$_2$ correspond to (101) and (001) planes of the rutile phase of TiO$_2$, respectively.[14] In SAED pattern, the bright and sharp diffraction spots of rutile TiO$_2$ (JCPDF 65-0192) are clarified, reflecting the single-crystalline nature of the H-TiO$_2$ nanowrie with (001) orientation.[15] The H-TiO$_2$@Ni(OH)$_2$ core-shell nanostructures are also unambiguously confirmed by the energy dispersive X-ray spectrometry (EDS) mapping analysis (Figure 1e), confirming that the inner NW core is TiO$_2$ and the outer-layer shell is Ni(OH)$_2$.

The desired morphology of Ni(OH)$_2$ nanosheets observed on the scaffold of H-TiO$_2$, indicates that the hydrogenation treatment produces a distinct surface condition (e.g., chemical activity/microstructure) of TiO$_2$, which could affect the subsequent nucleation and growth of Ni(OH)$_2$. Indeed, **Figure 2a** shows the O 1s core-level XPS spectra of TiO$_2$ and H-TiO$_2$ arrays. The black dots represent the experimental data, which can be deconvoluted into two peaks located at ~530 eV (blue dashed line) and 531-532 eV (pink dashed line) respectively. The peak centred in the lower binding energy range corresponds to the characteristic peak of O atoms in Ti-O-Ti.[16] Additional peaks centred at 531.3 and 532.2 eV are attributed to O atoms in Ti-OH, which have been reported to be ~1.5-1.8 eV higher than the case of Ti-O-Ti.[17] The relative intensity of Ti-OH peak in the H-TiO$_2$ sample is substantially higher than that of TiO$_2$ sample, showing the TiO$_2$ surface is functionalized with hydroxyl groups after hydrogenation with H plasma. Figure 2b shows the normalized Ti 2p core level XPS spectra of TiO$_2$ and H-TiO$_2$ samples. The two broaden peaks located at ~458 eV and ~465 eV, present in both samples, correspond to the Ti 2p$_{1/2}$ and Ti 2p$_{3/2}$ states of the Ti$^{4+}$, respectively. However, there is a clear shift toward lower energy in H-TiO$_2$ compared with TiO$_2$, indicating different oxidation states of Ti atoms. When subtracting the normalized Ti 2p spactra of H-TiO$_2$ with TiO$_2$, one can see two extra peaks located at 459.5 and 465.5 eV, which can be assigned to the Ti 2p$_{1/2}$ and Ti 2p$_{3/2}$ peaks of Ti$^{3+}$ and therefore confirming the presence of Ti$^{3+}$ ions in the H-TiO$_2$ samples.[17,18]

We have further performed first-principles calculations within the framework of density functional theory (DFT) to uncover the origin of chemical activity in TiO$_2$ affected by hydrogenation process. We find that a strong charge transfer occurs between the adsorbed H and TiO$_2$ surface, thus creating a charged layer in the proximity of the surface which facilitates the decoration of the Ni(OH)$_2$ layer over TiO$_2$ surface. There are four types of atomic species on the TiO$_2$ (110) surface (Figure 2c): the threefold coordinated in-plane oxygen (O$_{3c}$), the twofold coordinated bridging oxygen (O$_{2c}$), fivefold coordinated in-plane titanium (Ti$_{5c}$), and sixfold coordinated titanium (Ti$_{6c}$) located under the O$_{2c}$ atom. The most energetic favorable site for H adsorption is the O$_{2c}$ site with the adsorption energy ($E_a$) of -2.67 eV, indicating a strong chemical bonding of H-O$_{2c}$. The adsorption of H on O$_{3c}$ is weaker with $E_a$ of -2.07 eV while direct adsorptions of H on Ti$_{5c}$ and Ti$_{6c}$ are unstable. We then investigated the charge transfer for H adsorbed above the $O_{2c}$. Isosurface of differential charge density (DCD) $\Delta\rho(r)$ is shown in Figure 2d, where a strong charge depletion around H atom and a charge accumulation on the $TiO_2$ surface are observed, indicating that the anchored H atom serves as a n-type dopant. To accurately calculate the number of electrons donated to the surface, one can calculate the planar averaged DCD $\Delta\rho(z)$ along the normal z direction by integrating $\Delta\rho(r)$ within the (110) surface plane. The number of electrons transferred from H to the surface up to $z$ point is estimated by $\Delta Q(z)$. A strong oscillation of the $\Delta\rho(z)$ curve at the H-$TiO_2$ interface suggests that H induces a significant redistribution of surface charge. The maximum value of the profile (see the horizontal dashed line in Figure 2e) within the H-$TiO_2$ interface region gives the exact number of the donated electrons from H, amounting to 0.24 e per each H atom. As shown in Figure 2d, most of these excess electrons are localized around the $Ti_{6c}$ directly underneath the $O_{3c}$ and they effectively reduce the covalent state of Ti atom from $Ti^{4+}$ to $Ti^{(4-\delta)+}$, where $\delta$ is a small positive number. Accumulated electron cloud at the outer shell leads to an increase in the radius of Ti atom. More importantly, the excess electrons in $TiO_2$ mainly occupy the conduction band, and the population of these anti-bonding empty states weakens the Ti-O bond and triggers the deviation of Ti atom from the equilibrium position as shown by the green arrows in Figure 2f. With increasing exposure of the surface to H source, oxygen vacancy ($O_V$) is likely to form on the bridging site. Previous study has shown that $O_V$ can also donate excess electrons where the excess electron mainly distributes at the surface $Ti_{5c}$ atom.[19] Nevertheless, the built-in potential owing to the surface dipole is basically the same as the direct adsorption of H atom.

Both the XPS results and DFT caculations reveal the presence of the hydrogenation-induced charged defects ($[Ti^{3+}]_{Ti}^{4+}$), which are believed to contribute to a promoted surface activity of $TiO_2$ and consequently affect the growth of $Ni(OH)_2$. The formation of nanostructured shell materials was generally affected by the interface conditions including microstructure, lattice orientation or lattice mismatch at the nucleus-substrate interface boundary, which cause nucleation on foreign body with different free energy.[20] In addition to these crystallographic features, the introduction of surface charges (e.g., holes or defects) is also able to control the growth of deposited-materials, as the electrostatic is a powerful force for selectively adhesion of charged particles onto surfaces with polarity.[13] In other words, the localized surface charges were supposed to work as active cites to anchor nuclei, which subsequently grow and yield a desired surface morphology of the nanostructures. Herein, for the H-TiO$_2$ surface, the hydrogenation effectively modifies the surface chemical activity of TiO$_2$ NWs by introducing Ti$^{3+}$ species. These negative charged sites of the [Ti$^{3+}$]$_{Ti}^{4+}$ defects tend to absorb the positive-charged ions of Ni$^{2+}$ through electrostatic interaction. Compared to TiO$_2$, H-TiO$_2$ possesses more active sites to anchor the Ni(OH)$_2$ seeding, enabling the subsequent radom-growth of Ni(OH)$_2$ with ultrathin sheets.

The electrochemical behaviour of H-TiO$_2$@Ni(OH)$_2$ architecture was characterized in a three-electrode cell in 6M KOH aqueous electrolyte using Pt as counter electrode and Hg/HgO as reference electrode. **Figure 3a** compares the cyclic voltammogram (CV) collected for H-TiO$_2$@Ni(OH)$_2$ and TiO$_2$@Ni(OH)$_2$ at a scan rate of 5 mV s$^{-1}$ with potential window ranging from -0.1 to 0.4 V. Apparent redox peaks are observed in the CV curves of the H-TiO$_2$@Ni(OH)$_2$ and TiO$_2$@Ni(OH)$_2$, which are the characteristics of typical pseudocapacitance with the reversible reactions of Ni$^{2+}$/Ni$^{3+}$.[21] The current density of the H-TiO$_2$@Ni(OH)$_2$ electrode is higher than that of TiO$_2$@Ni(OH)$_2$, convincingly suggesting an improved function of H-TiO$_2$ NWs for the growth of electrochemically active Ni(OH)$_2$. Interestingly, the potential separation between anodic and cathodic peaks is narrowed in CV signals of H-TiO$_2$@Ni(OH)$_2$, as compared to that of TiO$_2$@Ni(OH)$_2$. This is commonly related to the kinetic limitation, which is caused by ionic or electronic hindrance. The narrowed potential window therefore implies an improved conductance when H-TiO$_2$ is used as the scaffold. The CV curves of the H-TiO$_2$@Ni(OH)$_2$ electrode at different scan rates (Figure S4a), show a small potential shift between anodic and cathodic peaks, which also demonstrates a good electrical conductivity. Given that the $TiO_2$-based NW electrode has negligible contribution to capacitance compared to that of $Ni(OH)_2$-based electrode, the specific capacities of H-$TiO_2$@$Ni(OH)_2$ and $TiO_2$@$Ni(OH)_2$ were calculated based on the mass of $Ni(OH)_2$ (Figure 3b). The H-$TiO_2$@$Ni(OH)_2$ yields an greatly improved capacitive performance with the maximum capacity reaching as high as 306 mAh g$^{-1}$, which is nearly 2 times higher than that of the $TiO_2$@$Ni(OH)_2$ at the same scan rate of 1 mV s$^{-1}$. Moreover, with increasing scan rate up to 100 mV s$^{-1}$, a retention value of 65% (193 mAh g$^{-1}$) is still maintained, which is greatly higher than that of 3% (5.2 mAh g$^{-1}$) for $TiO_2$@$Ni(OH)_2$. This clearly indicates that the improved capacitive retention arises from the effect of H-$TiO_2$ NWs. In addition, the galvanostatic charge/discharge measurements are conducted to examine the electrochemical performance of H-$TiO_2$@$Ni(OH)_2$ and $TiO_2$@$Ni(OH)_2$ (Figure 3c, Figure S4b). The charge/discharge curve of the H-$TiO_2$@$Ni(OH)_2$ is substantially prolonged over $TiO_2$@$Ni(OH)_2$, revealing an excellent capacitive behaviour.

Electrochemical impedance spectroscopy (EIS) measurements of H-$TiO_2$@$Ni(OH)_2$ and $TiO_2$@$Ni(OH)_2$ were conducted (Figure 3d). The impedance spectra of the H-$TiO_2$@$Ni(OH)_2$ and $TiO_2$@$Ni(OH)_2$ exhibit similar semicircles at high frequency region and straight lines in the low frequency region. One notes that the spike-like region in H-$TiO_2$@$Ni(OH)_2$ is closer to the imaginary axis, which demonstrates fast charge-transfer kinetics and electric responses resembling a circuit with low resistance and large capacitance connected in parallel.[22] In contrast, the impedance curve of the pristine $TiO_2$@$Ni(OH)_2$ is far from the imaginary axis indicating mass diffusion obstacles that the ion or charge transfer is experienced in the electrode. In addition, the serial resistance (intercept on real axis) and charge transfer resistance (diameter of the semicircle) of the H-$TiO_2$@$Ni(OH)_2$ is much lower than that of the $TiO_2$@$Ni(OH)_2$. It has been reported that the serial resistances are composed of the inter-granular electronic resistance (between active particles) and the contact resistance (between active materials and current collector), while the charge transfer resistance is associated with the electrode/electrolyte interface.[23] The EIS data thus demonstrates that the H-TiO$_2$@Ni(OH)$_2$ presents a much enhanced conductivity in both the serial and charge transfer parts, which is contributed by the tailored configuration of Ni(OH)$_2$ and the high conductivity of H-TiO$_2$. The efficient charge transport from the Ni(OH)$_2$ to carbon cloth via H-TiO$_2$ NW core also explains the improved electrochemical performance of H-TiO$_2$@Ni(OH)$_2$ electrodes.

Asymmetric supercapacitor (ASC) devices comprising of different materials for two electrodes allow for a broad operating voltage window, giving rise to improved energy densities. Before assembling an asymmetric full cell device, we have fabricated N-doped carbon NWs grown on carbon cloth (N-C NWs) using as negative electrode, which were derived through carbonizing PPys NWs (Figure S5). It demonstrates a maximum capacitance of 303 F g$^{-1}$ can be achieved at a scanning rate of 5 mV s$^{-1}$ and high specific capacitance values of 142 F g$^{-1}$ can be maintained at higher scanning rate of 200 mVs$^{-1}$. An asymmetric full cell device is fabricated based on H-TiO$_2$@Ni(OH)$_2$ as positive electrode and N-C NWs as negative electrode. The mass ratio of the positive electrode to negative electrode active materials is controlled at around 1:4. **Figure 4a** shows the CV curves collected at different voltage windows for the H-TiO$_2$@Ni(OH)$_2$//N-C NWs ASC at the scan rate of 10 mV s$^{-1}$ in 6M KOH electrolyte. Unlike the sharp redox peaks observed in the CV curve of the Ni(OH)$_2$@H-TiO$_2$ electrode, the asymmetric supercapacitor shows a rectangle-like CV curve in a large potential range from 0 to 1.8 V, indicating a typical capacitor behavior of electrochemical double layer capacitors (EDLCs). This behavior has been observed in previous studies on Ni(OH)$_2$@CNT//AC-based ASC.[7] The cell voltage of the H-TiO$_2$@Ni(OH)$_2$//N-C ASC cell can reach up to 1.8 V, which is almost twice as that of conventional carbon-based supercapacitors in aqueous electrolytes (~1 V).[24] This is a critical factor in determining the energy density, considering that the cell energy density is given by $E = 1/2\ CV^2$, where $C$ represents the specific capacitance and $V$ is the cell voltage. As shown in Figure 4b, the discharge curves of the ASC device present a good linear relation of discharge/charge voltage versus time, as an indication of the capacitor-like behavior.[25] The specific capacitances, on the basis of the galvanostatic curves, are calculated using the mass of $Ni(OH)_2$ and N-C NWs (Figure 4c). The maximum specific capacitance is 150.6 F $g^{-1}$ at 1 A $g^{-1}$, which is much higher to the values reported recently for the $Co_3O_4$@$Ni(OH)_2$//graphene (110 F $g^{-1}$),[25] $Ni(OH)_2$/graphene//porous graphene (~125 F $g^{-1}$),[24] $Ni(OH)_2$/CNT//AC (~120 F $g^{-1}$),[7] α-$Ni(OH)_2$//AC (125 F $g^{-1}$).[26]

A durable and stable cycling life of the supercapacitors is highly important for applications.[17] In this work, the long-term cycling performance of the ASC devices was evaluated in the voltage window of 0-1.8 V at the discharge current density of 1 A $g^{-1}$ for 5000 cycles (**Figure 4d**). The ASC device exhibits excellent stability with 90% retention of the initial capacitance after 5000 cycles. Such cycling performance is highly competitive with those of some other ASC devices, such as $Ni(OH)_2$//AC (82% retention after 1000 cycles),[27] CNT@$Ni(OH)_2$//AC (83% retention after 3000 cylces),[7] $Ni(OH)_2$//AC (82% retention after 1000 cycles),[26] NiO//carbon (ca. 50% retention after 1000 cycles),[28] $LiNi_{1/3}Co_{1/3}$//AC (~80% retention after 1000 cycles),[29] graphene/$MnO_2$//graphene (79% retention after 1000 cycles).[30] With respect to the power density and energy density, being important descriptors of the performance of supercapacitor devices, Figure 4e compares the specific power density(P) and energy density (W) of the as-fabricated ASC devices with the previously reported nickel-based ASC in aqueous electrolyte solutions.[31] In the present work, the H-$TiO_2$@$Ni(OH)_2$//N-C NWs ASC device with a cell voltage of 1.8 V exhibits an energy of 70.9 Wh $kg^{-1}$ at a power density of 102.9 W $kg^{-1}$, and still can retain 13.5 Wh $kg^{-1}$ at a power density of 20.9 k W $kg^{-1}$. Interestingly, we note that the *P* and *W* values are higher than the $TiO_2$@$Ni(OH)_2$//AC devices, suggesting the positive roles of H-$TiO_2$ nano arrays. Moreover, the maximum energy density (70.9 Wh $kg^{-1}$) is substantially higher than those reported for NiO- or $Ni(OH)_2$-based ASCs, such as $Ni(OH)_2$/graphene//$RuO_2$/graphene (48 Wh $kg^{-1}$)[31]

NiO//carbon (15-20 Wh kg$^{-1}$)[27] Ni-Zn-Co oxide/hydroxide//carbon (41.65 Wh kg$^{-1}$),[32] Co$_3$O$_4$@Ni(OH)$_2$//AC (41.83 Wh kg$^{-1}$),[25] CNT@Ni(OH)$_2$//AC (50.6 Wh kg-1),[7] Ni(OH)$_2$//AC (42.3 Wh kg$^{-1}$).[26] We also note that our energy density not only approaches the highest value (77.8 Wh kg$^{-1}$) in the Ni(OH)$_2$/graphene//porous graphene, but also shows a better retention ability.[24] As shown in Figure 4e, at a power density of ~1.7 kW kg$^{-1}$ as an example, an energy density of 56.9 Wh kg$^{-1}$ (retention 81% of maximum value) is obtained in H-TiO$_2$@Ni(OH)$_2$//N-C NWs, while only 20 Wh kg$^{-1}$ (retention of 26% of maximum value) can be remained in Ni(OH)$_2$/graphene//porous graphene.[24] In addition, we also assembled a solid-state asymmetric supercapacitor in a full cell set up consisting of the N-C NWs as the negative electrode and H-TiO$_2$@Ni(OH)$_2$ as the positive electrode in gel KOH/PVA electrolyte. The CV curves collected for the ASC device under flat, bent, and twisted conditions are almost overlapped (Figure 5f), indicating their excellent mechanical stability for flexible energy storage devices.

In summary, we demonstrate a new approach to control the morphologies and structures of core-shell hybrids through judiciously engineering the interface with intentially introduced defects. With pre-chemical functionalization of TiO$_2$ surface via hydrogenation, an electrochemically favourable porous Ni(OH)$_2$ with ultrathin nanosheets are anchored on H-TiO$_2$ NW arrays. This as-designed core-shell hierarchical nanostructure shows an improved specific capacitance of as high as 306 mAh g$^{-1}$, which is almost 2 times that of the counterpart in TiO$_2$@Ni(OH)$_2$. As proof-of-concept application in devices with high energy and power densities, an asymmetric supercapacitor with H-TiO$_2$@Ni(OH)$_2$ core-shell NWs as the positive electrode and the N-C NWs as the negative electrode was fabricated. The obtained device can operate in a 1.8 V voltage window and deliver a high specific capacitance of 150.6 F g$^{-1}$ with a maximum energy density of 70.9 Wh kg$^{-1}$, as well as a stable cycling ability of 90% up to 5000 cycles and a good structural flexibility. The novel surface-charge-mediated method in this study sheds some light on controllable synthesis of arbitrary core-shell structures for applications in high-performance electrochemical energy storage.

**Experimental Section**

*Preparation of self-supported H-TiO$_2$@Ni(OH)$_2$ core-shell NWss:* In this work, the self-supported TiO$_2$ NWs arrays on carbon cloth were prepared based on the modification of a previously reported hydrothermal method.[33] Then the TiO$_2$ NWs was placed in vacuum chamber and irradiated with hydrogen plasma at 450 ºC for 5000 s to generate H-TiO$_2$ NWs. To grow Ni(OH)$_2$ shells on the H-TiO$_2$ NWs, a solution was prepared by mixing 16 mL of 1M NiSO$_4$·6H$_2$O, 20 mL of 0.25 M K$_2$S$_2$O$_4$ and 4 ml of aqueous ammonia (24% NH$_3$·H$_2$O) in a Pyrex beaker at room temperature.[34] The H-TiO$_2$ NWs were immersed into the aqueous solution for 10 min to derive H-TiO$_2$@Ni(OH)$_2$. NWs. As a comparison, TiO$_2$@Ni(OH)$_2$ NWs were also prepared under same conditions, but without hydrogenation process.

*Fabrication of Aqueous ASC:* Aqueous ASC was assembled by H-TiO$_2$@Ni(OH)$_2$ and N-C NWs electrodes with a separator (normal filter paper). The weight ratio of H-TiO$_2$@Ni(OH)$_2$ positive electrode to N-C NWs negative electrode was fixed to be 1:4. 6 M KOH aqueous solution was used as the electrolyte.

*Fabrication of Solid-State ASCs:* The solid-state ASC devices were assembled by separating the H-TiO$_2$@Ni(OH)$_2$ and N-C NWs electrodes with normal filter paper sandwiched in between. PVA/KOH gel was used as electrolyte and prepared by mixing 4.2 g of PVA and 4.2 g of KOH in 50 ml of deionized water, and heated at 40 °C for 12 h under vigorous stirring. Before assembling, one piece of H-TiO$_2$@Ni(OH)$_2$ (1 cm×2cm) and two pieces of N-C NWs (1 cm×2cm) were soaked with PVA/KOH. The flexible solid-state supercapacitor was prepared after the gel electrolyte solidified at room temperature.

*Material Characterization and Electrochemical Measurement:* The morphology and structure of electrode materials were characterized by SEM (SEM, Zeiss), TEM (TEM, JEOL 2010), XRD (Bruker AXS, D8 Advance, Cu Kα, λ = 0.154060 nm), XPS (Kratos Analytical).

Electrochemical performance was evaluated by cyclic voltammetry (CV), galvanostatic charge-discharge and impedance spectroscopy by using Solartron System 1470E and 1400A, respectively. The electrochemical impedance measurement was biased to 0 V (vs. the open-circuit voltage) using the three-electrode system, at a frequency ranging from 0.001 to 100 kHz with a 5 mV RMS voltage perturbation amplitude.

*Computational methods:* The $TiO_2$ (110) surface is modeled by creating a periodic slab with twelve atomic layers together with a vacuum of 15 Å thickness normal to the surface. Atoms in the three lowest atomic layers are fixed in their bulk positions. We build a (3 × 2) surface supercell (three unit cells in the [001] direction and two in the [1$\bar{1}$0] direction), where the lattice constants of the unit cell are $\sqrt{2}a$ and $c$ along [1$\bar{1}$0] and [001] directions, respectively. The a = 4.57 Å and c = 2.94 Å are relaxed lattice constants of bulk $TiO_2$.[35] Spin-polarized first-principles calculations are performed by using Vienna ab initio simulation package (VASP) within the framework of DFT.[36] The Perdew-Burke-Ernzerhof functional and the projector augmented wave method (PAW-PBE) are adopted. A cutoff energy of 400 eV and the 4 × 3 × 1 Monkhorst-Pack (MP) grid for k-point sampling are used. All the structures are relaxed until the Hellmann-Feynman forces on each atom become less than 0.01 eV/Å. The adsorption energy $E_a$ is defined as $E_a = E(H+TiO_2) - E(H) - E(TiO_2)$, where $E(H+TiO_2)$, $E(H)$ and $E(TiO_2)$ are the energies of the H adsorbed $TiO_2$, H atom, and pristine $TiO_2$ surface, respectively. The differential charge density $\Delta\rho$ is calculated via $\Delta\rho = \rho(H+TiO_2) - \rho(H) - \rho(TiO_2)$, where $\rho(H+TiO_2)$ is the electron density of H adsorbed $TiO_2$, and $\rho(H)$ and $\rho(TiO_2)$ are the electron densities of H atom and $TiO_2$ surface in the positions of the adsorbed system, respectively.

**Supporting Information**

Supporting Information is available from the Wiley Online Library or from the author.

**Acknowledgements**

The author thanks the financial support provided by MOE, Singapore Ministry of Education (Tier 2, MOE2012-T2-2-102), for research conducted at the National University of Singapore. This work was supported by MOE under AcRF Tier 2 (ARC 26/13, No. MOE2013-T2-1-034; ARC 19/15, No. MOE2014-T2-2-093; MOE2015-T2-2-057) and AcRF Tier 1 (RG5/13), and NTU under Start-Up Grant (M4081296.070.500000) in Singapore. The authors gratefully acknowledge the use of computing resources at the A*STAR Computational Resource Centre, Singapore.

Received: ((will be filled in by the editorial staff))
Revised: ((will be filled in by the editorial staff))
Published online: ((will be filled in by the editorial staff))

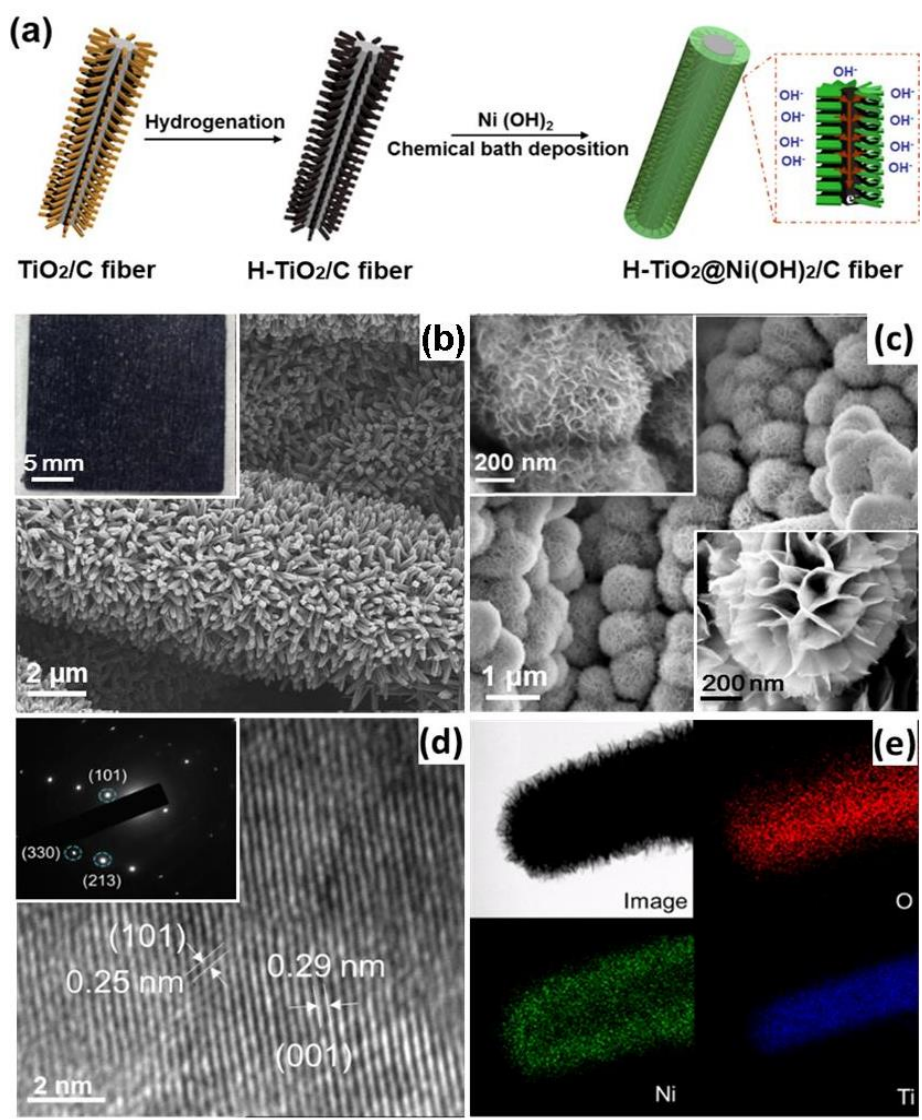

**Figure 1.** a) Schematic illustration for the fabrication of H-TiO$_2$@Ni(OH)$_2$ core-shell NWs arrays. b) SEM images of H-TiO$_2$ NWs arrays. The inset shows the optical image of H-TiO$_2$ NWs. c) SEM image of H-TiO$_2$@Ni(OH)$_2$ NWs, the magnified structures of H-TiO$_2$@Ni(OH)$_2$ and TiO$_2$@Ni(OH)$_2$ are shown in the inset of upper and lower corners respectively. d) HRTEM images of H-TiO$_2$ NW. Inset is the SAED pattern of the H-TiO$_2$ NW. e) EDS mapping results for a single hybrid nanowire, demonstrating the TiO$_2$ core and Ni(OH)$_2$ shell hierarchical structure.

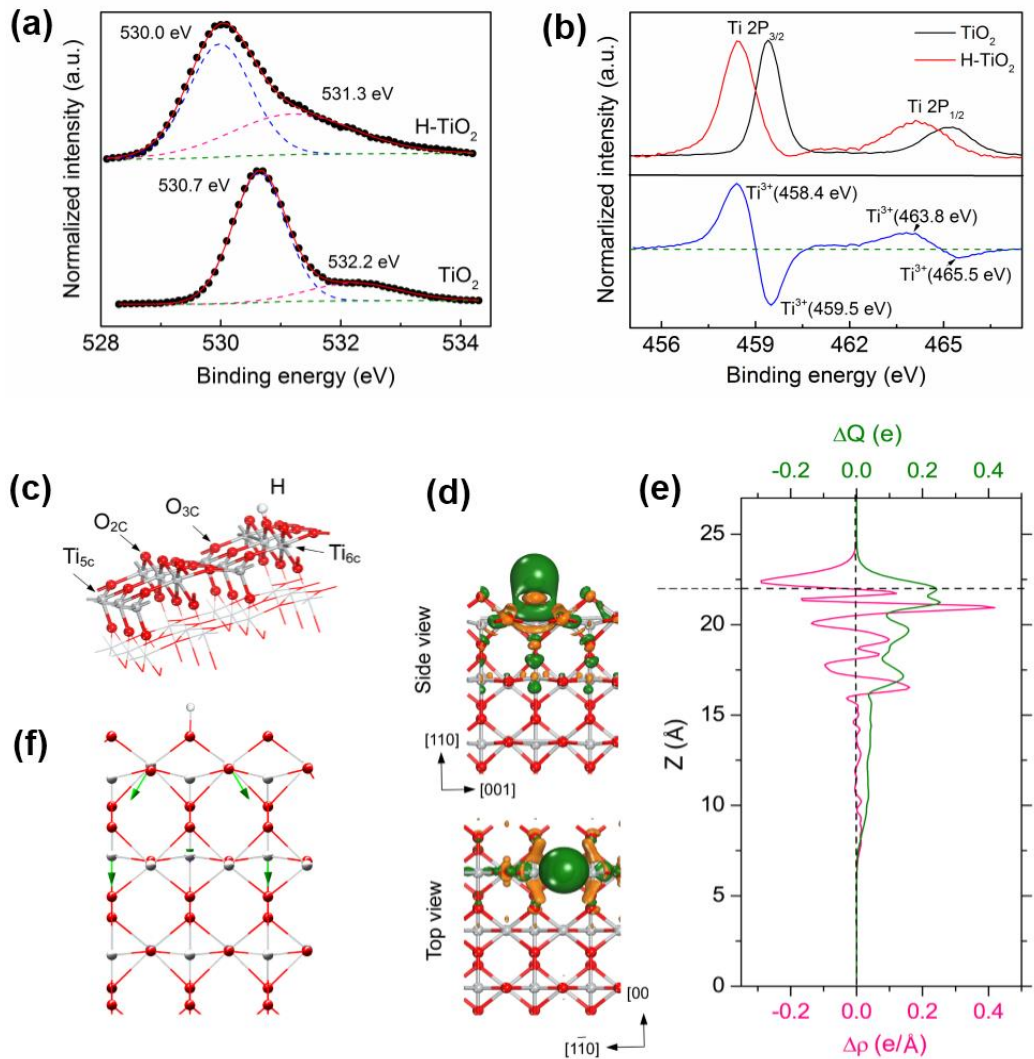

**Figure 2.** Surface chemical activity affected by hydrogenation studied by XPS and First-principle caculations. a) Normalized O 1s core level XPS spectra of pristine $TiO_2$ and $H-TiO_2$. b) Overlay of normalized Ti 2p core level XPS spectra of $TiO_2$ and $H-TiO_2$, together with the spectrum of their differences (spectra of $H-TiO_2$ minus that of $TiO_2$). c) Atomic model of the $TiO_2$ (110) surface with H atom adsorbed on the bridging oxygen site. d) Isosurface of differential charge density $\Delta\rho(r)$ where the violet (green) color denotes depletion (accumulation) of electrons. (e) Profiles of the plane-averaged differential charge density $\Delta\rho(z)$ (red line) and the $\Delta Q(z)$ (green line). f) Atomic displacements upon H adsorption, where significant deviation from the equilibrium position is observed for $Ti_{6c}$ atoms and the length of the green arrows is proportional to the magnitude of the atomic adjustment.

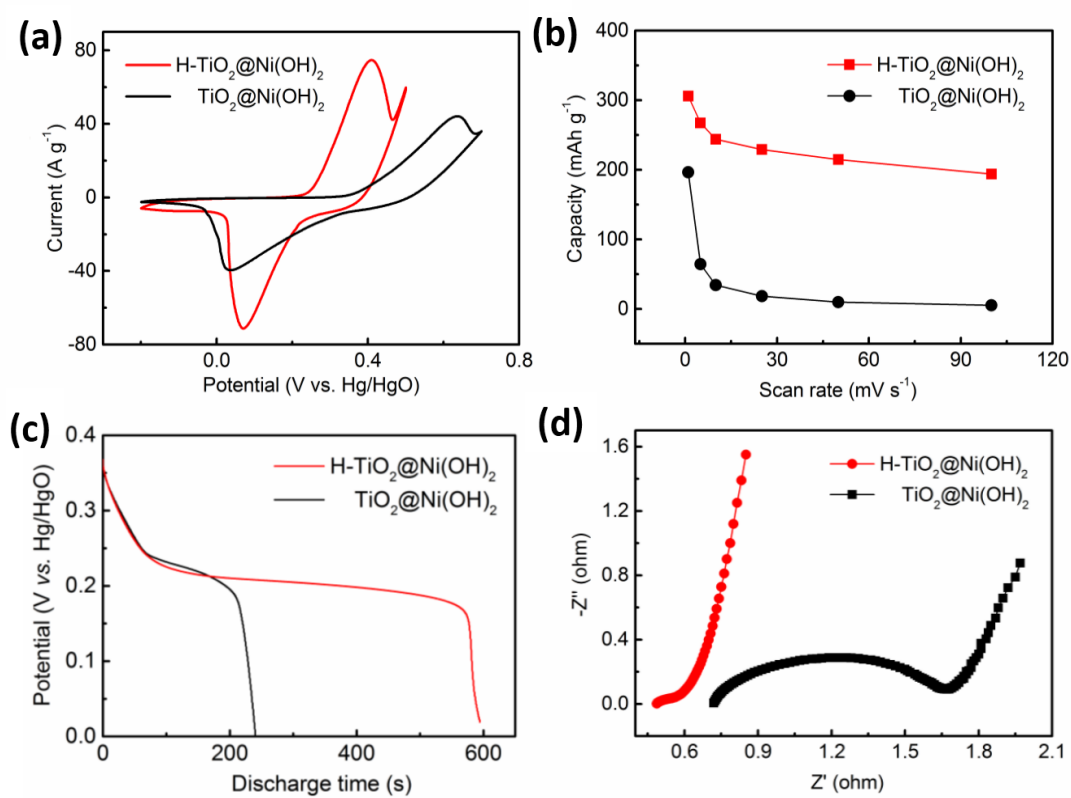

**Figure 3.** Comparison of electrochemical performance of H-TiO$_2$@Ni(OH)$_2$ and TiO$_2$@Ni(OH)$_2$ in a three-electrode configuration. a) CV curves measured at a scan rate of 5 mV s$^{-1}$. b) Capacity at different scan rates. c) Galvanostatic discharge curves collected at a discharge rate of 1 A g$^{-1}$. d) Impedance Nyquist plots collected with 6 M KOH electrolyte.

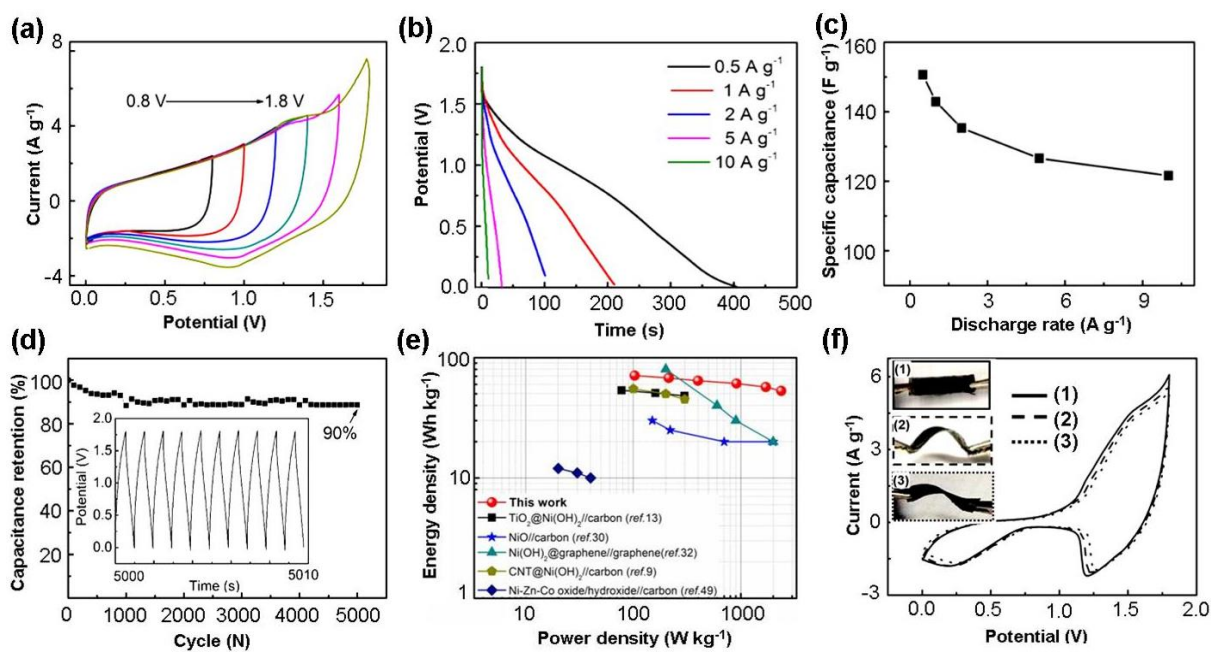

**Figure 4.** Electrochemical performance of ASC device in the aqueous electrolyte (6M KOH). a) CV curves of the ASC device collected in different scan voltage windows. b) Galvanostatic charge/discharge curves collected at different current densities for the ASC device operated within a voltage window of 1.8 V. c) The specific capacitances calculate from the Galvanostatic curves as a function of scan rate. d) Cycling performance of ASC devices collected at discharge rate of 1 A g$^{-1}$ for 5000 cycles. The inset shows followed 10 cycles of the charge-discharge curves after 5000 cycles. e) Ragone plots of the ASC devices measured in the aqueous electrolyte of 6M KOH. The values reported for other NiO- and Ni(OH)$_2$-based devices are also added for comparison. f) Flexible solid-state ASC device is assembled based on H-TiO$_2$@Ni(OH)$_2$ as the positive electrode and N-C NWs as the negative electrode in gel KOH/PVA electrolyte and CV curves collected at a scan rate of 25 mV s$^{-1}$ for the ASC device under flat, bend and twisted conditions. Insets are the device pictures under test conditions.

**An electrochemically favorable Ni(OH)$_2$ morphology with porously hierarchical structure and ultrathin nanosheets** is achieved through modulating the surface chemical activity of TiO$_2$ by hydrogenation, which creates an defect-rich surface of TiO$_2$, thereby facilitating the subsequent nucleation and growth of Ni(OH)$_2$. This configuration-tailored H-TiO$_2$@Ni(OH)$_2$ core-shell Nanowires is shown an improved electrochemical performance.

**Keyword**

supercapacitors, hydrogenation, asymmetric, nanowires, surface charge

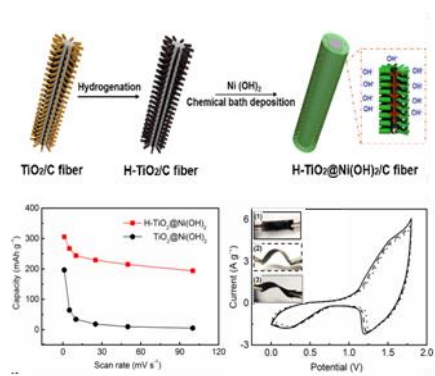

# Supporting Information

for Adv. Mater., DOI: 10.1002/adma.((please add manuscript number))

**Surface-charge-mediated Formation of H-TiO$_2$@Ni(OH)$_2$ Heterostructures for High-Performance Supercapcitors**

*Qingqing Ke, Cao Guan, Xiao Zhang, Minrui Zheng, Yongwei Zhang, Yongqing Cai,\* Hua Zhang,\* John Wang\**

**Supporting results**

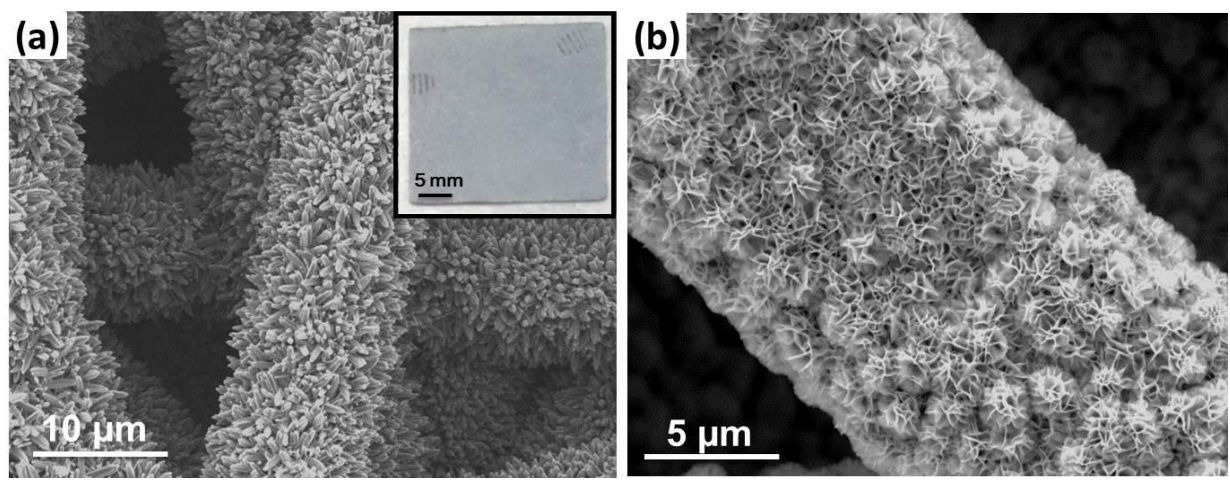

**Figure S1.** a) SEM image of TiO$_2$ NWs arrays. The inset shows the optical image of TiO$_2$ NWs. b) SEM image of TiO$_2$@Ni(OH)$_2$ NWs.

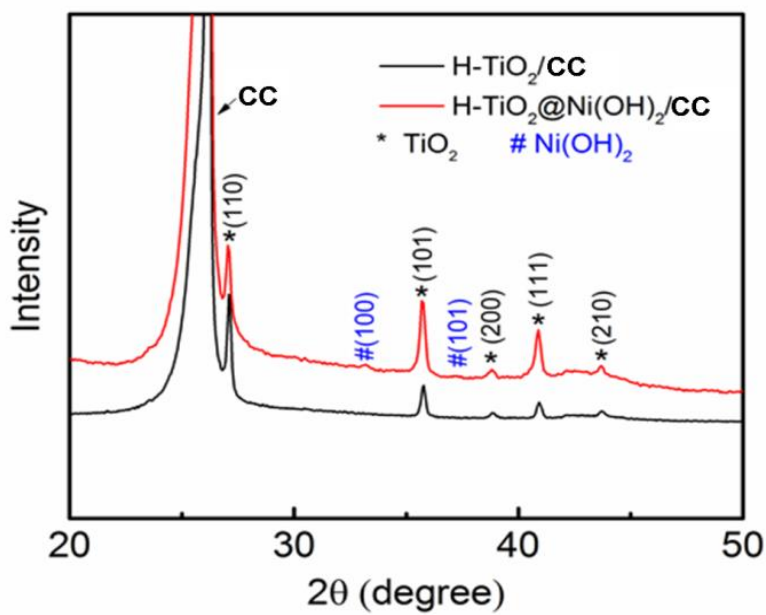

**Figure S2.** XRD patterns of H-TiO$_2$ (black line) and H-TiO$_2$@Ni(OH)$_2$. The diffraction pattern of the H-TiO$_2$ is also collected as a reference for comparison. The diffraction peaks positioned at 2θ of 27.9º, 36.6º, 40.01º, 41.7 º, 44.4º assigning to (110), (101), (200), (111) and (210) respectively, confirm the rutile phase of the TiO$_2$ NWs. In addition to the diffraction

peaks from TiO$_2$, peaks located at 33.7º and 38.4º were observed, which are assigned to (100) and (101) of β-Ni(OH)$_2$ (JCPDS card no. 1-1047).[1]

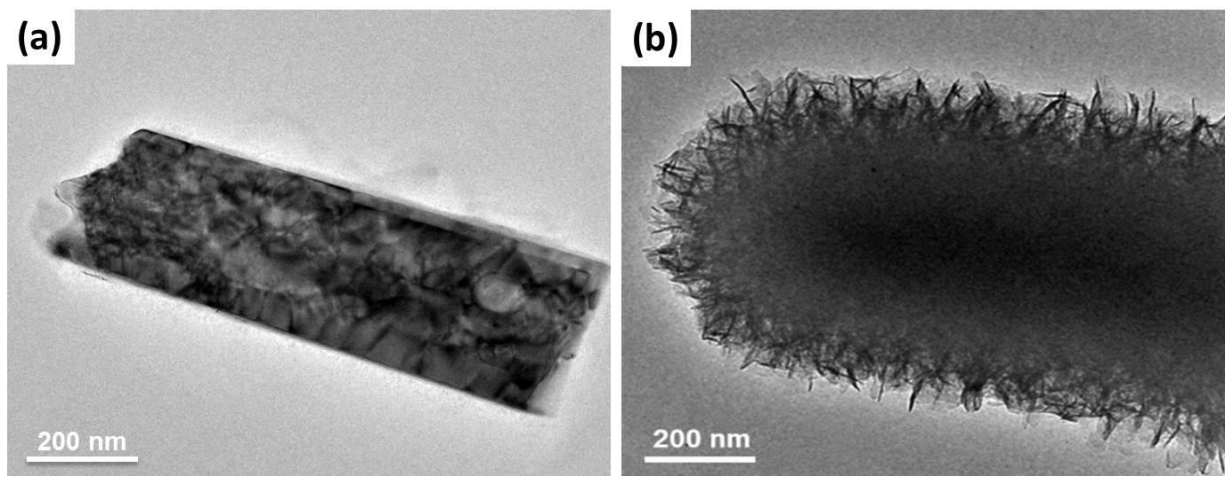

**Figure S3:** TEM images of H-TiO$_2$ NW and H-TiO$_2$@Ni(OH)$_2$ NW

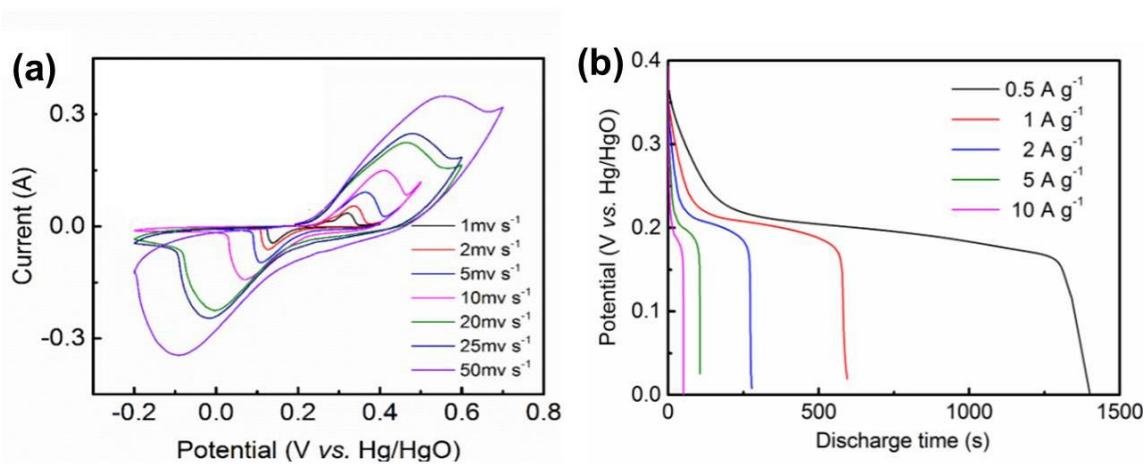

**Figure S4:** a) Electrochemical performance of H-TiO$_2$@Ni(OH)$_2$. b) Charging-discharging curves for H-TiO$_2$@Ni(OH)$_2$ at different discharge rates. The CV curves of the H-TiO$_2$@Ni(OH)$_2$ electrode at different scan rates, in order to evaluate faradaic processes. A small potential shift between anodic and cathodic peaks is observed, demonstrating a good electrical conductivity, which could well benefit high-rate energy storage in pseudocapacitors. The charge-discharge behaviour of the H-TiO$_2$@Ni(OH)$_2$ electrode is determined between 0 and 0.5 V at different current densities. At each current density, the electrode charges and discharges rapidly with good electrochemical reversibility.

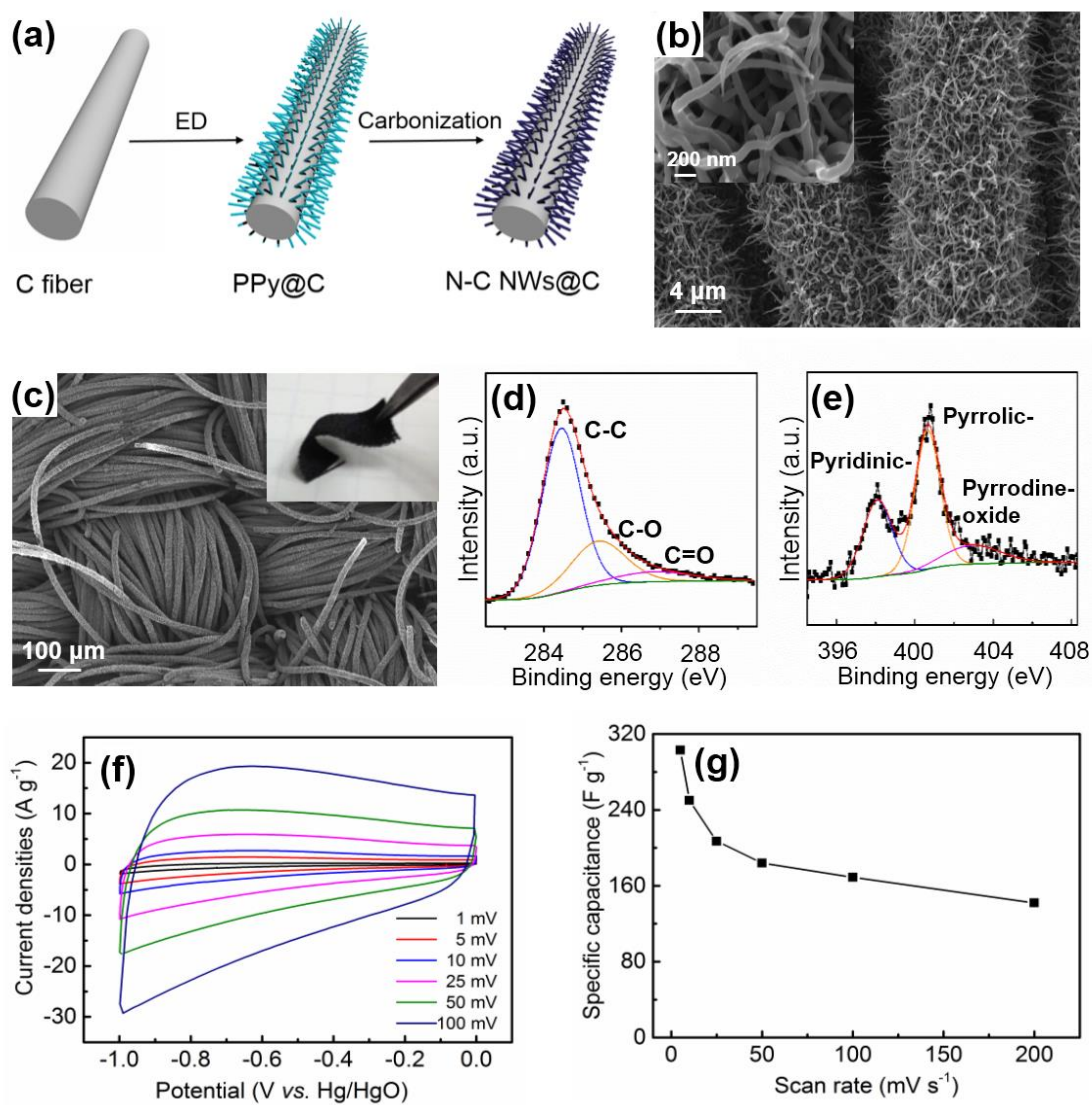

**Figure S5.** a) Schematic illustration for the development of N-C NWs. b) and c) SEM images of N-C NWs at various magnifications and inset of (c) shows the flexibility of N-C NWs. d) and e) XPS spectra of C and N respectively. f) CV curves measured at various scan rates. g) The corresponding specific capacitances with respect to the different scanning rates.

To fulfil the high-power characteristic of an asymmetric supercapacitor, carbon-based materials such as active carbons, graphene and CNTs are generally employed as negative electrode.[2,3] N-doped carbon materials have been recently attracted great attention due to their high electrical conductivity, which provides higher charge/electron carrier mobility.[4,5] In this work, we have fabricated N-doped carbon NWs grown on carbon cloth (N-C NWs), which were derived through carbonizing PPys NWs (Figure S5a, see experiment part for details). The N-C NWs with average diameter of ~100 nm are vertically aligned on the carbon fiber surfaces (Figure S5b). The textile structure of the carbon cloth is preserved after the growth of N-C NWs. This leads to an excellent mechanical stability to permit the potential application for flexible energy storage devices (Figure S5c). The XPS results suggests that PPys was effectively reduced after annealing at 800 °C, which is closely related to the enhanced conductivity of the carbon NWs. The presence of N dopants in the carbon NWs is also clarified in XPS spectrum (Figure S5e). There are three types of N atoms in carbon with different characteristics binding energies: pyridinic-N (398.5 eV), pyrrolic-N (400.5 eV), and pyridine-N-oxide (403 eV).[6] It has been know that nitrogen doping can act as an electron-donor during the charge-discharge process, therefore, they are likely to provide higher charge/electron carrier mobility to enhance the electrical conductivity of carbon NWs.[5] The CV curve of N-C NWs measured in 6 M KOH solution exhibits a typical rectangular shape without obvious distortion even at a scan rate of 200 mV s$^{-1}$ (Figure S5f), indicating an excellent capacitance behavior and fast diffusion of electrolyte ions into the electrode. The specific capacitance of the N-C NWs at different scanning rates is summarized in Figure S5g. A maximum capacitance of 303 F g$^{-1}$ can be achieved at a scanning rate of 5 mV s$^{-1}$, Moreover, we also note that the as-derived electrode material not only exhibits high specific capacitance values but also gives rise to a high value of 142 F g$^{-1}$ at higher scanning rate of 200 mVs$^{-1}$.

*Experimental section for preparation of N-doped carbon NWs grown on carbon cloth (N-C NWs):* Firstly, PPy NWs arrays were typically prepared by electrochemical polymerization on carbon cloth.[7] In brief, carbon cloth was firstly coated with Au by sputtering with a current of 40 mA for 10s. The electrochemical polymerization solution (200 mL) contains disodium hydrogen phosphate dodecathydrate (15.406 g, Na$_2$HPO$_4$·12H$_2$O), sodium phosphate monobasic dehydrate (6.240 g, $NaH_2PO_4 \cdot 2H_2O$), p-toluenesulfonyl (3.884 g, 98.5%,) and pyrrole (1.388 mL, 98%,). The as-prepared carbon cloth coated with Au was used as the work electrode. A Pt plate (2×2 $cm^2$) and a saturated calomel electrode (SCE) were used as the counter and reference electrode, respectively. The electrochemical experiments were carried out by using an electrochemical workstation (Solartron System 1470E). To electrochemically deposit PPy on carbon cloth, the typical current applied to the working electrode was 2 mA and kept for 4 h. In the followed step, the prepared PPy NWs is carbonized by annealing at 800 °C in $N_2$ atmosphere for 2h in order to well derive the N-C NWs.